\documentclass{emulateapj}

\newcommand\colaa {\null}
\newcommand\cola {&}
\newcommand\colb {&}
\newcommand\colc {&}
\newcommand\cold {&}
\newcommand\cole {&}
\newcommand\colf {&}
\newcommand\colg {&}
\newcommand\colh {&}
\newcommand\coli {&}
\newcommand\colj {&}
\newcommand\colk {&}
\newcommand\eol{\\}

\shorttitle{Star Clusters in NGC 4038/39}
\shortauthors{Whitmore et al.}

\def\msun{M$_{\odot}$}
\def\arcsec{{\tt ''}}			
\def\etal{et al.}			
\def\H0{$H_0$~= 75 \kms\ Mpc$^{-1}$}

\def\halpha{H$\alpha$}

\def\kms{km s$^{-1}$}

\def\ref{\par\noindent\hangindent 30pt}

\def\farcs{\hbox{$.\!\!^{\prime\prime}$}}

\def\v16{$\Delta V_{1-6}$}

\begin{document}

\title{STIS Spectroscopy of Young Star Clusters in ``The Antennae'' Galaxies
(NGC 4038/4039)\footnote{
Based on observations with the NASA/ESA {\it Hubble Space 
Telescope}, obtained at the Space Telescope Science Institute, which is
operated by the Association of Universities for Research in Astronomy,
Inc., under NASA contract NAS5-26555.
}}

\author{Bradley C. Whitmore, Diane Gilmore,
C. Leitherer, S. Michael Fall, and Rupali Chandar}
\affil{Space Telescope Science Institute,
Baltimore, MD 21218}
\email{whitmore, dkarakla, leitherer, fall, rupali@stsci.edu}

\author{William P. Blair}
\affil{Johns Hopkins University, Department of Physics \& Astronomy, 
Baltimore, MD 21218}
\email{wpb@pha.jhu.edu}

\author{Fran\c cois Schweizer}
\affil{Carnegie Observatories, 813 Santa Barbara Street, Pasadena, CA
91101-1292}
\email{schweizer@ociw.edu}

\author{Qing Zhang}
\affil{Cadence Design Systems Inc., San Jose,
CA 95134}
\email{qzhang2b@yahoo.com}

\and

\author{Bryan W. Miller}
\affil{AURA/Gemini Observatories, La Serena, Chile}
\email{bmiller@gemini.edu}

\begin{abstract}

Long-slit spectra of several dozen young star clusters have been
obtained at three positions in the Antennae galaxies with the Space
Telescope Imaging Spectrograph (STIS) and its $52\arcsec\!\times
0\farcs2$ slit.  Based on \halpha\/ emission-line measurements, the
average cluster-to-cluster velocity dispersion in seven different
cluster aggregates (``knots'') is $<$10 \kms.  The fact that this
upper limit is similar to the velocity dispersion of gas in the disks
of typical spiral galaxies suggests that the triggering mechanism for
the formation of young massive compact clusters (``super star
clusters'') is unlikely to be high velocity cloud--cloud collisions.
On the other hand, models where preexisting giant molecular clouds in
the disks of spiral galaxies are triggered into cluster formation are
compatible with the observed low velocity dispersions.  These
conclusions are consistent with those reached by Zhang et al.\ (2001)
based on comparisons between the positions of the clusters and the
velocity and density structure of the nearby interstellar medium.  We
find evidence for systematically lower values of the line ratios
[N~II]/H$\alpha$ and [S~II]/H$\alpha$ in the bright central regions of
some of the knots, relative to their outer regions. This suggests that
the harder ionizing photons are used up in the regions nearest the
clusters, and the diffuse ionized gas farther out is photoionized by
`leakage' of the leftover low-energy photons.  The low values of the
[S~II]/H$\alpha$ line ratio, typically [S~II]/H$\alpha$ $<$ 0.4,
indicates that the emission regions are photoionized rather than shock
heated.  The absence of evidence for shock-heated gas is an additional
indication that high velocity cloud--cloud collisions are not playing
a major role in the formation of the young clusters.

\end{abstract}


\section{Introduction}

The discovery of young massive star clusters in merging galaxies was
one of the important early results from the {\it Hubble Space
Telescope} (Holtzman et al.\ 1992; Whitmore et al.\ 1993).  These clusters
have all the attributes of young globular clusters, hence providing us with
the opportunity to study the formation of these systems in the local
universe.  Dozens of other galaxies have now been found to harbor similar
clusters.  Merging galaxies appear to have the largest populations of
such ``super star clusters,'' though similar clusters are also
found in smaller numbers in starburst, barred, and even normal spiral
galaxies (see Whitmore 2003 for a review). This discovery has acted as
a catalyst for the study of star clusters and for the study of star
formation in general since it now appears that most stars form in
clusters (e.g., Lada \& Lada 2003; Fall et al.\ 2005).

What triggers the formation of these young, compact, massive clusters?
One possibility is that high-velocity collisions of the gaseous
components in interacting galaxies result in the formation of
large-scale shock fronts that trigger gravitational collapse and the
subsequent formation of stars and clusters (e.g., Gunn 1980; Schweizer 1987; 
Kang et al.\ 1990; Olson \& Kwan 1990a, 1990b; Noguchi 1991;
Kumai et al.\ 1993a, 1993b).
Specifically, Kumai et al.\ 1993a suggest that chaotic cloud--cloud
collisions with velocities $\geq$ 50 -- 100 \kms\ may result in the formation of young
globular clusters due to local shock compression in the Magellanic Clouds.
A more recent model by Bekki \etal\/ (2004) suggests that velocities in the
range 10 -- 50 \kms\/ are optimal for making clusters.
Recent numerical, smooth-particle-hydrodynamic simulations of NGC 4676
(``The Mice'') by Barnes (2004) have been successful in reproducing
the extended nature of star formation in this interacting galaxy pair
with a shock-induced star-formation law.  However, it is important to
note that much of the extended star formation in this model originates in
relatively low-velocity shocks due to convergent flows, rather than in
high-velocity shocks created by interpenetrating collisions of
galactic disks.
Some observational support for cluster formation being triggered by
high-velocity collisions comes from Wilson et al.\ (2000), who find that
the strongest CO and IR source in the Antennae galaxies coincides
with a knot in the ``Overlap Region'', where two giant molecular clouds
appear to be colliding with differential velocities of 50 -- 100 \kms.

There is, however, direct evidence against high-speed
cloud collisions being the dominant triggering mechanism for 
the formation of star clusters in the Antennae galaxies. 
Zhang et al.\ (2001) made detailed comparisons between the 
positions of star clusters in the Antennae and the stellar 
and interstellar contents of these galaxies, based on 
observations over a wide range of wavelengths. They found 
little or no correlation between the positions of young 
clusters (very near their birthplaces) and the velocity 
structure of the surrounding interstellar medium, as
measured by gradients and dispersions in the HI and 
H$\alpha$  channel maps at various distances from the 
clusters, as would be expected if cloud motions triggered 
the formation of the clusters (see Figures 15, 16, and
17 of Zhang et al.\ 2001). In contrast, Zhang et al.\ found 
a strong correlation between the positions of the young
clusters and several other properties, including
gaseous emission from young stars
and the density structure of the surrounding interstellar 
medium as measured by the HI and CO intensity maps. The
implications of these results for the formation of the clusters are
discussed further by Fall (2004).

In the present paper we measure the velocities of several dozen young
clusters in the Antennae by using the high spatial resolution made
possible by the {\it Hubble Space Telescope} ({\it HST\,}), and from
these compute the corresponding velocity dispersion.  While previous
ground-based studies (e.g., Burbidge \& Burbidge 1966; Rubin et al.\
1970; Amram et al.\ 1992) mapped out the velocities of various
``knots'' in The Antennae (e.g., labeled A through T by Rubin et al.),
the Hubble observations by Whitmore \& Schweizer (1995) revealed that
each of these ``knots'' consists of roughly a dozen or more individual
star clusters.  Long-slit spectroscopic observations with the Space
Telescope Imaging Spectrograph (STIS) can provide a direct measurement
of the velocity dispersion between clusters within such knots.
If clusters formed in high-speed cloud collisions, we might expect the resulting 
velocities of the clusters, and hence their dispersion, to 
be large. The main caveat here concerns the possibility 
that some of the kinetic energy of cloud motions is 
dissipated and radiated away during the collisions and hence 
is not fully reflected in the velocity dispersion among 
the clusters. If we neglect this dissipation for the 
moment, the observed velocity dispersion among the 
clusters should reflect the velocity dispersion of the
clouds from which they formed. This is the test we 
perform here, as a complement to the one reported by 
Zhang et al. (2001). 

Throughout the present paper
we adopt a Hubble Constant of $H_0 = 75$ km s$^{-1}$ Mpc$^{-1}$, which
for a heliocentric systemic velocity of $cz_{\odot} = 1630$ \kms\
(corresponding to 1440 \kms\ relative to the Local Group)
places NGC 4038/4039 at a distance of 19.2 Mpc.  At this distance, the
projected scale is $1\arcsec = 93$~pc, and the corresponding distance
modulus is 31.41 mag.                  
We note that Saviane et al.\ (2004) determine a distance of only
$13.8 \pm 1.7$~Mpc for The Antennae, based on the tip of the red giant
branch observed in the southern tidal-tail population.  This distance
would imply a peculiar velocity of $\sim$400 \kms,
which would be quite high for a field galaxy in a loose group such as
NGC 4038/4039. We prefer to use the original redshift distance for this
paper, pending a more definitive measurement of the photometric distance.

\section{Observations and Reductions}

Long-slit spectra of the Antennae galaxies were obtained with STIS during
the period from 2000 April to 2001 March, as part of Proposal GO-8170.
A variety of observations were made, including long-slit H$\alpha$
observations (G750M grating), both long-slit and slitless observations in
the UV (G140L grating), and slitless observations in the blue (G430M
grating). Most results in the current paper are based on the H$\alpha$
observations.
More detailed results from observations
with the G140L and G430M  gratings will be reported in a future paper.
Chandar \etal\/ (2005) also  use the G140L spectra
in their study of the ages of clusters in 27 star-forming galaxies. 
Their spectroscopically determined ages are in good agreement with
the photometrically determined ages reported in the present paper.

Table 1 presents a variety of measured and derived parameters for the
cluster candidates discussed in this paper. These include the
cluster ID number (from Whitmore \& Zhang 2002), a quality
index based on how well the \halpha\ and I-band images correspond, absolute
magnitude ($M_V$), age estimate (from Whitmore \& Zhang 2002), equivalent
width of \halpha\/, velocity, line width, relative \halpha\ flux, and [NII]/H$\alpha$\/ and [SII]/H$\alpha$\/ line strength
ratios. As discussed below, 
in cases when the emission appears to be diffuse (e.g., the quality = 3 candidates), we include the nearest cluster
in the table in order to obtain information about the physical
characteristics within the knot (e.g., the mean age of nearby clusters).

A $52\arcsec\!\times 0\farcs2$ slit was used with the G750M grating to
cover the spectral region of $\lambda\lambda$6500 -- 7050 \AA, hence
including the [NII] $\lambda\lambda$6548,\,6583, H$\alpha$, and [SII]
$\lambda\lambda$6716,\,6731 lines. The 
scale was 0.05$\arcsec$ pixel$^{-1}$, the reciprocal spectral dispersion 
0.56 \AA\ pixel$^{-1}$, and the spectral resolution for a nominal point 
source $\sim$0.9 \AA\ (FWHM). 

For the observations,
three slit positions were selected to include as many young clusters as
possible, as shown in Figure 1. We note the curious linear alignments
of clusters utilized in visits 6 and 7. It is unclear whether these
alignments are due to chance, the edges of occulting dust clouds (which
are often found to be filamentary and roughly linear), or perhaps 
shock fronts formed at the intersection of the two colliding galactic
disks.  In any case, the fortuitous alignments allowed us to
include one to two dozen clusters along each of the three slit positions.

Figures 2a--c show the three long-slit spectra. Note that the continuum
can only be seen for the brightest knots. Figure 3 shows an interesting,
peculiar steep-gradient feature
in the spectrum of Knot B (Visit 6). This
feature is visible in both the \halpha\ and [NII] lines. 
The fact that the feature appears tilted indicates that it stems
from a spatially directed, rather than  spherically symmetric
outflow.
There is no obvious corresponding feature visible in the  
\halpha\ and broad-band
images.

The three exact positions of the slit were determined post facto by comparing
the \halpha\ flux profiles measured along the slit with linear 
brightness extractions
from our \halpha\ image (Whitmore et al.\ 1999) taken with the
Wide-Field Planetary Camera 2 (WFPC2). 
Figures 4a--c compare the imaging and spectral profiles for each visit.
Using this method we were able to
determine positions of the slit with an accuracy of about $0\farcs1$,
good enough to deduce which specific objects fell on the slit (Table 1).
Note that the slightly poorer spatial resolution of the
WFPC2 H$\alpha$ image relative to the STIS data makes some of the objects
appear less prominent in Figs.\ 4a--c. 
Figures 5a--c show the slit positions
superposed on both \halpha\ and $I$-band WFPC2 images.

Note that not all of the peaks in the \halpha\/ profiles are due to
clusters; some are due to shells and other  
diffuse emission between the clusters.  In particular,
the fraction of H$\alpha$ peaks associated with diffuse emission rather than
clusters was higher in Knot  T, presumably because the clusters there tend
to be slightly older and have had more time
to expel their gas.  We include a ``quality'' index
to represent the degree of certainty of our cluster assignments 
in column 4 of Table 1. This is also
why we refer to ``candidate'' clusters rather than one-to-one matches.
In cases when the emission appears to be diffuse, we include information about the  nearest cluster
in the table in order to obtain information about the physical
characteristics within the knot (e.g., the mean age of the
nearby clusters). Hence it should be kept in mind that for the quality = 3
objects, the various properties
(e.g., the equivalent widths) listed in Table 1 are not for the same region that
we have sampled from our slit spectrum.

Table 2 summarizes the mean properties of the seven primary knots
covered by our slit positions. The ages of the clusters within these
knots were derived using the technique described in Whitmore \& Zhang
(2002), but using the Bruzual \& Charlot (2003) models rather than
their 2000 models. Note the correlation between the quality index
and the age of the clusters. As expected, the younger clusters show a
better correspondense between their H$\alpha$ and continuum
morphologies (i.e., lower quality index), since the gas has not had
much time to be spatially displaced from the clusters. We also note the extreme
youth ($\sim$2 Myr) of some of these clusters, befitting their
large equivalent widths.

The limited pointing accuracy, and difficulties in determining optimal
positions for the slit, resulted in typical deviations of a few
tenths of an arcsecond from some desired positions.
This caused us to miss some of the brightest clusters
(e.g., we skirted the edge of Knot B in Visit 7). However, with so
many clusters in each knot, missing one cluster generally meant that
we picked up another cluster somewhere else along the slit. Hence the total
number of clusters was roughly the same as originally planned.

The original data were recalibrated using CALSTIS with the most recent
reference files. Hot pixels were removed with an IDL routine that
replaces deviant pixels by the median of the adjacent pixels.  Line
strengths, line widths (FWHM of a Gaussian profile), and velocities
were measured using the SPECFIT program within STSDAS, with results given
in Table~1. The simplex algorithm was used to interactively
fit Gaussians to all five emission lines simultaneously.  Measurements
were made both in and between the clusters whenever the line flux
warranted it. Extraction regions ranged from 3 pixels ($0\farcs15$) for
bright regions to 24 pixels ($1\farcs2$) where the line flux was
weak. Values for the velocity were taken from the \halpha\ peak
alone since the weaker lines are absent for faint sources.
Figures 6a--c present the relative \halpha\ flux (solid lines),
velocity $cz$, \halpha\ line width, [NII]/\halpha\ line ratio, and
[SII]/\halpha\ line ratio for the three visits.

A potential complication  is the fact that the clusters 
are partially resolved within the $0\farcs2$-wide slit. This means
that part of the measured velocity is due to the off-center position
of clusters within the slit. Because of this, the  
velocity
dispersions reported below represent upper limits.
In principle it should be possible to correct for this effect
by convolving  the \halpha\ image with the slit transmission function. 
However, a rough empirical check indicated that this effect was barely
discernible, whence no corrections were applied for the present paper.
The affect is very small, in any case, as demonstrated by the small
RMS deviations of radial velocities along the slit, as we will discuss
later in the paper.

This complication is more serious for our G140L and G430M
spectra, most of which were taken with a $0\farcs5$ rather than a
$0\farcs2$ slit. We attempted to circumvent this problem by  
measuring wavelengths for the Milky Way components of strong 
interstellar lines 
(e.g., C~II~$\lambda1335$ and Si~II~$\lambda1260$) in order to 
determine a wavelength zeropoint
for the spectra.  Due to the radial velocity of the Antennae, ISM
lines arising in the Antennae are redshifted relative to the same
lines arising in the Milky Way; hence, in principle we should be able
to see both the Galactic and Antennae components. Figures 21 and 23
in Whitmore \etal\/ (1999) show that this is indeed the case for
Knots S and K, that are dominated by two of the brightest clusters in the Antennae.
Unfortunately, the S/N  and spatial resolution 
of the current spectra were too low  
for this method to be applied here successfully.

A comparison of the newly determined velocities with published values 
for Knots B, C, D, E, F,
and T shows a relatively large range in absolute differences, especially
against the early low-dispersion  work by Burbidge \& Burbidge (1966; mean difference of
+48 \kms, with a dispersion of 62 \kms).
The agreement is much better for velocities measured by 
Rubin et al.\ (1970; mean difference of +10 \kms,
dispersion of 34 \kms) and Amram et al.\ (1992; mean difference of +15
\kms, dispersion of 18 \kms).  A comparison with the
HI velocities from Hibbard et al.\ (2001) is problematic,
due to the large mismatch in spatial resolution  (i.e., 
$\sim$$0\farcs1$ for the
HST data and $\sim$10$\arcsec$\/ for the HI data), but a rough estimate
yields a mean difference of --3 \kms\/ and a dispersion of 48 \kms. 
The differences are in the sense
that the earlier papers found larger velocities.

\section{Results}

\subsection{Cluster-to-Cluster \halpha\ Velocity Dispersions}

A primary goal of this project is to determine whether high
cloud--cloud velocity dispersions are the triggering mechanism
for the formation of super star clusters, as predicted by some models
(e.g., Kumai et al.\ 1993a, Bekki \etal\/ 2004).  

Assuming the gas clouds were already present before the collision (i.e.,
GMCs), one might expect that the resulting dispersion after the
encounter would be comparable to the original velocity difference
between the two colliding systems.
For example, if portions of two colliding galaxies encountered each other
with relative velocities of 100 \kms, one might expect
a cluster forming from the collision of a high-mass cloud of
velocity 100 \kms\ with a low-mass cloud of velocity
0 \kms\ to have a resulting velocity near 100 \kms. On the other
hand, if instead  the high-mass cloud
had a velocity of 0 \kms, the resulting cluster would have
a velocity near 0 \kms. In practice the situation is likely to
be much more complicated,  with dissipation, multiple collisions, 
and turbulence all playing important roles. Hence, we would
expect the observed cluster-to-cluster velocity dispersion to be somewhat lower
than the original cloud--cloud velocity dispersion. 

A cursory glance at Figures 6a--c reveals immediately that the
velocity fields along the slit positions are relatively quiescent,
with velocity dispersions $\sim$10 \kms\ rather than the $\sim$100 
\kms\ difference in the orbital velocities of the two colliding
galaxies. {\it We note that these low observed cluster-to-cluster
velocity dispersions are similar to those found for gas, young
stars, and young clusters in normal spiral disks (Mihalas \& Binney 1981).}
While it is conceivable that the inelasticity of cloud--cloud
collisions discussed in the previous paragraph might reduce the
observed velocity dispersion of a system of clusters formed by high
velocity encounters to this degree, it seems too coincidental that the
resulting velocity dispersion would essentially match the value found
in normal disks. 

The low cluster-to-cluster velocity dispersions we have observed are
more compatible with models where the GMCs in the disks of the
colliding spiral galaxies are the seeds that form the young massive
clusters (e.g., Jog \& Solomon 1992; Elmegreen \& Efremov 1997),
rather than with models where large cloud--cloud velocity dispersions
are required (e.g., Kumai et al.\ 1993a, Bekki \etal\/ 2004). In the
former case the cluster-to-cluster velocity dispersion would be
roughly the same as the original cloud--cloud velocity dispersion,
i.e., similar to the velocity dispersion of the gas in a spiral disk.

These conclusions are consistent with those reached by Zhang
et al.\ (2001), as discussed in the introduction, who found little or no 
correlation between HI and H$\alpha$ velocity dispersions and the 
presence of young clusters in The Antennae.  
This interpretation is also supported by the fact that young massive
clusters are distributed throughout The Antennae rather than being
located only in the ``Overlap Region,'' where high-speed collisions
are most likely to occur.  
Similarly, young massive clusters are also found
in quiescent environments such as normal spiral galaxies
(Larsen \& Richtler 2000), albeit in smaller numbers. 

We have measured the cluster-to-cluster
velocity dispersion by selecting knots with three or more measured clusters
(i.e., Knots B, C, D, E, F, T, and T2), as listed in Table 2.
In several cases there are  velocity gradients
across the  knots which are part of the large-scale
velocity field (e.g., Knot F in Visit 7). We corrected for such gradients
by fitting a line to the velocities as a function of position in the knot
and computing the RMS of the velocity residuals.
The clusters in a typical knot extend $\sim$5$''$ along the slit, corresponding to
 $\sim$500 pc (see Table 2 for individual knots).

A potential complication for this procedure is that in some cases
(primarily Knot C) the velocity field is dominated by a super bubble
surrounding the knot, with the central part of the knot appearing
blue-shifted by $\sim$50 \kms\ with respect to its surroundings.  This
effect can be seen most clearly in Figure 2 of Amram et al.\ (1992),
where the velocity contours around Knots B and C appear to be
concentric rings. 

Table 2 shows that the cluster-to-cluster velocity dispersions measured
from the ionized gas are low in all knots.
Averaging over all seven knots yields a mean value for the
cluster-to-cluster velocity dispersion of 9.6 \kms.
Excluding Knot C from the average, because of the potential
bias introduced by the super bubble, yields a negligible difference (i.e.,
an average value of 9.5 \kms). 
As discussed in \S 2, the measured velocity dispersions represent upper
limits, since no correction has been made for the position of each cluster
within the slit.  

Globular cluster systems in galaxies show much 
larger velocity dispersions than gas in disks.  If the young massive
clusters in The Antennae are to form a cluster system with kinematics
similar to the kinematics of globular clusters in typical early-type
galaxies, they need to decouple from the gaseous kinematics.
This presumably happens later in the merger, when the process of
violent relaxation peaks.
The relatively ordered rotational velocity field observed in The
Antennae by Burbidge \& Burbidge (1966), Rubin et al.\ (1970; esp.\
\S IV), and Amram et al.\ (1992, esp.\ Fig.\ 2) argues that the
original disks are still relatively intact, and most of the violent
relaxation lies still ahead.

The low cluster-to-cluster velocity dispersions we have observed in
the Antennae beg the question of whether the multiple clusters in a
given ``knot'' are gravitationally bound to each other, and hence may
merge together to form more massive clusters, as advocated by various
authors (e.g., Kroupa 1998, Elmegreen \etal\/ 2000, Fellhauer
\etal\/ 2005, Bekki \etal\/ 2004). Using the virial theorem, and
typical values of M$_{knot}$ = 4 $\times$ 10$^6$ \msun\/ (each knot
typically including $\approx$ 25 individual clusters with mass greater
than 10$^4$ \msun\/) and R$_{knot}$ = 120 pc for our five best defined
knots (i.e., B, C, D, F, T), we estimate that the binding velocity is
$\approx$ 7 \kms\ (ranging from 5 to 10 \kms\ for the five knots). While
this is roughly at the limit where we can make reliable measurements
based on the current observations, we note that our observed upper
limit for the cluster-to-cluster velocity dispersion is only slightly
larger than this estimated binding velocity, hence the merging cluster
scenario may be viable, at least in part,  and deserves further study.

However, it is  important to note that the resulting structure that 
 forms
from the coalescence of {\it all} of the clusters in a knot would 
have a radius similar to the original radius of the knot
(i.e., $\approx$ 100 pc), and hence would look more like a dwarf
galaxy than a globular cluster. It may still be possible
to form structures with radii similar to globular clusters
(or the slightly larger ``faint fuzzy'' clusters observed by Brodie \& Larsen 2002)
if only two or three clusters within $\approx$ 10 pc of each other merge.

It is also interesting to compare the cluster-to-cluster \halpha\ velocity
dispersions with the HI velocity dispersions measured by Hibbard et
al.\ (2001) and here listed in the last column of Table 2.  The
measured HI velocity dispersions range from 15 to 30 \kms, typically a
factor of two higher than the cluster-to-cluster velocity dispersions.
This apparent discrepancy is almost certainly due to the poor spatial
resolution of the HI maps ($\sim$10$\arcsec$).  For example, Knots T
and F both show relatively large \halpha\ velocity gradients that we
remove when computing the cluster-to-cluster velocity dispersion, but
that remain included in the high measured HI velocity dispersions. If
the gradient is not removed the cluster-to-cluster \halpha\ velocity
dispersions are 28.9 \kms\ for Knot T and 31.2 \kms\ for Knot F;
very similar to the values for the HI velocity dispersions.

\subsection{\halpha\/ Line Widths}

In principle, it should be possible to learn about the kinematics of
the outflows around clusters and knots from the line-width
information.  Unfortunately, the ability to decipher this information
is compromised by the fact that the $0\farcs2$ width of the slit is
comparable to the width of the STIS point spread function.
Hence, an apparent increase
in the line width may be due to the presence of diffuse
line emission covering the slit more uniformly than a
point source would, rather than being indicative  of a kinematic signature.

According to Table 13.40 of the STIS Instrument
Handbook, the instrumental FWHM of spectral lines 
is $\sim$40 \kms\ for point sources and $\sim$105 \kms\ for very extended
sources. Note that the latter number is very approximate and likely
to be a slight underestimate, since
it assumes a simple rectangular profile without taking into account the
instrumental wings beyond the edges of the slit. In addition, the
program used for analysis (IRAF task SPECFIT)
fits a Gaussian profile rather than a rectangular profile. Hence,
direct comparisons with predictions of the line profile
for a uniform diffuse source are problematic.

The measured line widths in the high S/N knots range
from about 70 to 120 \kms\ (Figure 6). Hence, it seems likely that the observed
spread is due to the presence of a range of objects from partially
resolved compact star clusters to extended emission from gas that has
been expelled from the clusters and fills the slit nearly uniformly.
Because of the fact that the upper limit of the measured line
profiles is comparable to what is predicted for diffuse emission, we
conclude that in general there is no evidence for outflows with velocities
greater than $\approx$ 100 \kms.

Two possible exceptions occur in clusters \#7 and \#14 (Visit 6), both
with large apparent velocity widths ($>$150 \kms) and good signal-to-noise
ratios of \halpha\ (i.e., relative \halpha\ flux $>$50 in Table 1).  Cluster \#14 lies in the brightest
part of Knot B. Figure 3 shows a peculiar steep-gradient feature in its
spectrum, with a peak-to-peak velocity amplitude of 150 \kms.
This feature is probably responsible for the high velocity width measured
in Cluster \#14. The direct \halpha\ image (Fig.\ 5b) shows---to one side
of this cluster---the presence of a large shell-like structure that
may be associated with the spectral feature. The other high-velocity-width
feature is seen in Cluster \#7 (Visit 6), midway between Knots C and
D (Figs.\ 2b and 5b).  Careful inspection of Figure 2b and 6b suggests
that the superposition
of diffuse gas from Knots C and D, plus a discrete cluster at a slightly
lower velocity, may in this case be the explanation for the high apparent
velocity width.

Knot T represents a case where diffuse emission probably plays the
dominant role in determining the line width.  The correspondence
between the \halpha\ and $I$-band images is particularly bad in
this knot (i.e., ``quality'' rating of 3 for seven out of eight objects
in Table~1), probably because the clusters are older (mean age $\sim$6
Myr). Hence the gas has had more time to escape, resulting in 
diffuse emission which fills the slit more uniformly than a point source,  
resulting in higher apparent line widths
($\sim$110 \kms).

This interpretation is supported by the fact that the \halpha\ and
[N~II] lines across Knot T (Fig.\ 2a) show a series of wiggles, with
peak-to-peak velocity
amplitudes of $\sim$60 \kms\ (see Fig.\ 6a, around position 500).
Note from Fig.\ 6a that the line widths are $\sim$110 \kms, consistent
with diffuse emission filling the slit.  These wiggles are relatively
rare, and are largely responsible for Knot T having one of the
largest measured cluster-to-cluster velocity dispersions (14 \kms, see
Table 2).  Whitmore et al.\ (1999) inferred similarly high values for
the gas outflow velocities around Knots S and K based on the sizes of
the bubbles and age estimates.  Hence, it appears likely that the
observed line wiggles are related to the outflow velocities produced by
individual cluster winds around some of the older clusters.
In such cases, the {\it stellar} cluster-to-cluster velocity dispersions
are likely to be smaller than the gaseous velocity dispersion.
Even so, the gaseous velocity dispersion for Knot T is only 14 \kms.

The interpretation is simpler for knots that have low apparent
velocity widths, indicative of cases where the emission is coming from
compact regions around relatively point-like clusters rather than from diffuse emission that
fills the slit and artificially broadens the observed line width.  We
can infer that the clusters are particularly young in these cases and
have therefore not had time to blow sizable bubbles of diffuse
emission that fills the slit.  Probably the best example of this is Knot D (Fig.\ 6b),
with line widths around 80 \kms\/ and a relatively good correspondence
between the \halpha\ and $I$ images (i.e., no clear evidence for the
presence of shell-like structures such as those seen in Knots B and
C).  We also note that this is one of the younger knots, with a mean
age for the clusters of $\sim$2.5 Myr according to Table 2. In
addition, Fig.\ 5c shows that the spectrograph slit was positioned
near the center of this knot, whence most of the emission is coming
from the bright clusters rather than from the diffuse emission in the
area.  Similarly, Knot F has line widths around 80 \kms\/, and a mean
age of the clusters of $\sim$1.4 Myr.

We conclude that the apparent velocities derived from
line widths are primarily useful for
providing information about whether the emission is relatively
point-like or diffuse, rather than providing information about the
actual outflow velocities in the gas.  We note that this provides a
second-order method for estimating cluster ages, since the
knots with low apparent line widths (i.e., more point-like
emission) tend to contain the youngest clusters.

\subsection{Line-Strength Ratios}

Figures 6a--c show how the [N~II]/H$\alpha$ and [S~II]/H$\alpha$
line ratios change along each of the three long-slit positions.
In each case, what is plotted is the sum of two lines ($\lambda$6584
+ $\lambda$6548 for [N~II], $\lambda$6716 + $\lambda$6731 for [S~II])
relative to H$\alpha$.  

Several different effects are visible in these figures.  Overall, the
mean [N~II]/H$\alpha$ ratio observed is $\sim$0.4, and the mean
[S~II]/H$\alpha$ ratio is $\sim$0.2.  Faint knots, such as seen in the right half of
Fig.\ 6a (Visit 2 - Region T2), show a large dispersion due to observational
uncertainties (i.e., clusters with relative
\halpha\ flux $<$ 50 in Table 1 have observational uncertainties
$\sim$ 0.25), but are consistent with these mean
values.  Secondly, there is
evidence for systematically lower values of the line ratios in a few of the
knots
where the slit crosses emission of especially high surface brightness. This
results in the U-shaped line ratio patterns in Fig. 6b, Knots B, C,
and D.  This anti-correlation has been seen before in studies of
extragalactic H~II regions and diffuse ionized gas (DIG) (Hoopes et
al.\ 1999; Greenawalt et al.\ 1998).  Assuming that the DIG is
photoionized by `leakage' of leftover ionizing photons from
star-forming regions, the harder ionizing photons must be depleted
locally causing a higher mean ionization in the central parts of H~II
regions (and thus lower [S~II]/\halpha\ and [N~II]/\halpha\ ratios)
than in the surrounding, lower-surface-brightness diffuse gas.

The [S~II]/H$\alpha$ ratio is of interest because it has long been used as
a discriminant for shock-heated versus photoionized gas (see Blair \&
Long 1997; Gordon et al.\ 1998; and references therein).
Usually, a criterion with [S~II]/H$\alpha$ = 0.4 as the nominal threshold
is used.  Shock-heated nebulae tend to exhibit higher values of this
ratio (often considerably higher), while
photoionized nebulae show lower values (often below 0.2). We see no
bright, discrete nebulae sampled by the long slits here that exhibit
elevated [S~II]/H$\alpha$ ratios that would be indicative of shock
heating.  Hence, no supernova remnants containing bright radiative
shocks such as seen in the Cygnus Loop in our Galaxy (e.g., Fesen et
al.\ 1985) have been detected
serendipitously in these observations.  

The absence of obvious kinematic or line ratio evidence for shock-heated 
gas in the vicinity of the young clusters provides an
additional indication that high velocity cloud--cloud collisions are
not playing a major role in the formation of the young clusters.  We
should note, however, that our observations do not rule out the
presence of low--velocity shocks that heat and compress the molecular
clouds, since cooling takes place primarily through molecular lines
in the IR in this case, rather than leaving a signature in the
[S~II]/\halpha\/ ratio.

\section{Summary}

The Space Telescope Imaging Spectrograph (STIS) aboard the {\it
Hubble Space Telescope} has been used to obtain long-slit,
high-spatial-resolution spectra at three positions in the Antennae
galaxies.  Observations were made in three spectral regions: the
far UV (1150 -- 1720 \AA), the UV--Blue (3800 -- 4100 \AA), and the 
Red including the \halpha, [NII], and [SII] emission lines.
This paper focuses on the latter emission lines. The primary results
are as follows.

1. The average cluster-to-cluster \halpha\/ velocity dispersion in
seven cluster aggregates (``knots'') is $<$10 \kms\ and, therefore,
similar to the velocity dispersion of the gas and young stars 
in the disks of spiral
galaxies.

2. This low cluster-to-cluster velocity dispersion does not favor models that
require high-velocity cloud--cloud collisions to trigger cluster formation
(e.g., Olson \& Kwan 1990a,\,b; Kumai et al.\ 1993a, Bekki \etal\/ 2004).  On the other hand,
models
where giant molecular clouds (GMCs) already present in galactic disks act
as the seeds from which young star clusters form are compatible with the
observed low velocity dispersions.
This also supports earlier findings (Zhang et al.\ 2001) that
there is little or no correlation between
HI and H$\alpha$ velocity dispersions  and the
presence of young clusters in The Antennae.  

3. The line ratios of [N~II]/H$\alpha$ and [S~II]/H$\alpha$ across knots often
show U-shaped patterns, with low values where the spectrograph slit crosses
the brightest clusters. This suggests that the harder ionizing photons are
depleted in regions nearest the clusters, and the diffuse ionized gas farther
out is photoionized by `leakage' of the leftover low-energy photons.

4. The observed low values of the [S~II]/H$\alpha$ line ratio (typically
$\la$0.4) indicate that the HII regions associated with knots and
clusters are photoionized rather than shock heated.
No supernova remnants containing bright radiative shocks have been
detected serendipitously by our observations.  
The absence of evidence for shock-heated gas in the vicinity of the young clusters provides an
additional indication that high velocity cloud--cloud collisions are
not playing a major role in the formation of the young clusters. 

\acknowledgments
We thank Joshua Barnes for useful discussions, and the referee for
several suggestions that improved the paper.  This work was supported
by NASA grant GO-08710, and is based on observations with the NASA/ESA
{\it Hubble Space Telescope}, obtained at the Space Telescope Science
Institute, which is operated by the Association of Univeristies for
Research in Astronomy, Inc., under NASA contract NAS 5-26555.  One of
us (F.S.) gratefully acknowledges partial support from the NSF through
grant AST-02\,05994.

\begin{figure}[h]
\epsscale{.70}
\plotone{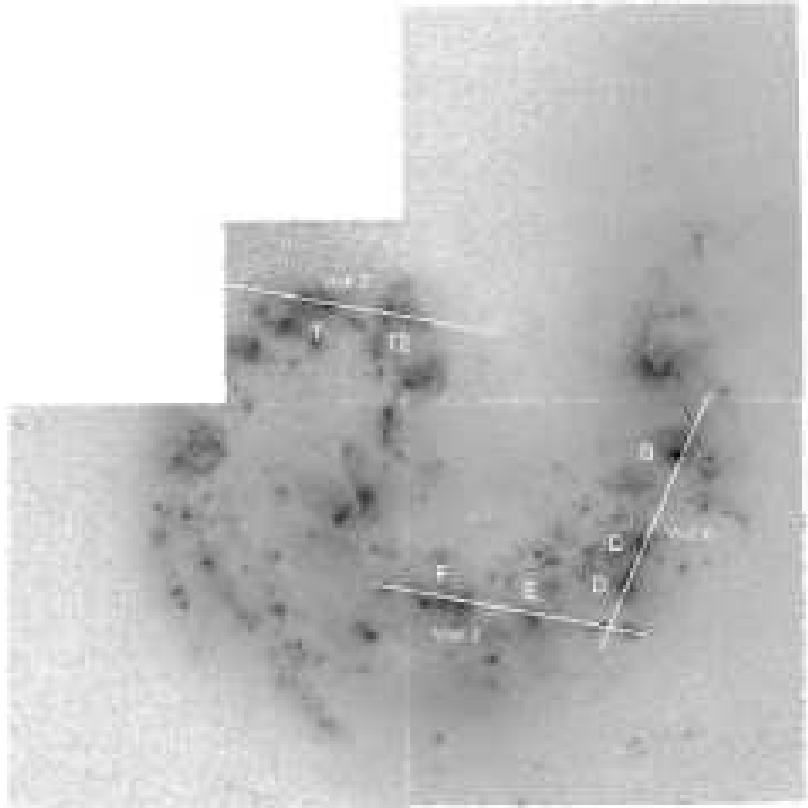}
\caption{Slit positions superposed on the $I$-band 
{\it HST\,} image of The Antennae for the three visits using the \halpha\ grating.
Capital letters identify knots named by Rubin et al.\ (1970).}
\label{fig1}
\end{figure}

\begin{figure}
\epsscale{.5}
\plotone{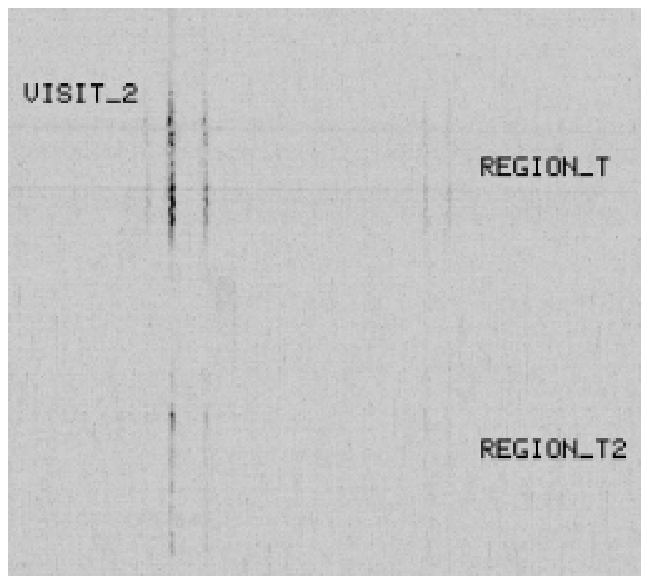}\\
\smallskip
\epsscale{1}
\plottwo{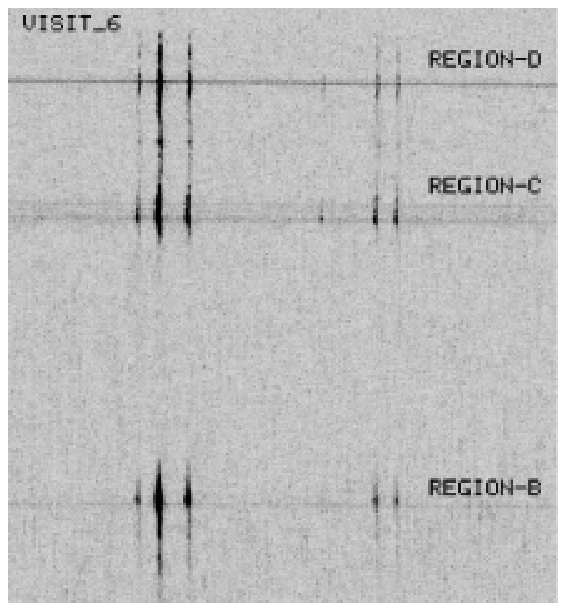}{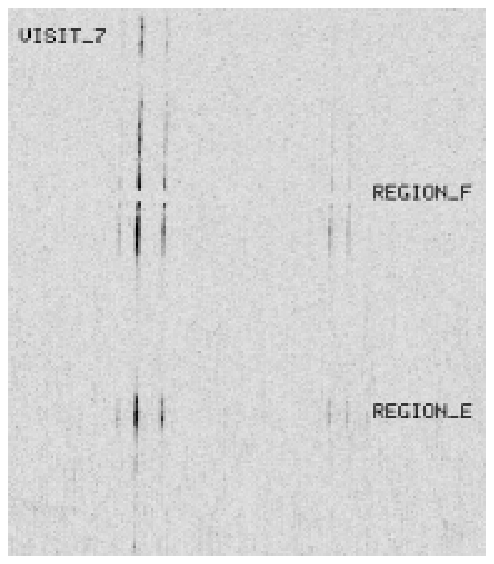}
\caption{(2a) Spectrum obtained with the \halpha\ grating during Visit 2. 
Wavelength increases toward the right; the spatial dimension is up and down.
The lines are (left to right): [NII] $\lambda$6548, H$\alpha$, [NII] 
$\lambda$6583, and [SII] $\lambda\lambda$6716, 6731. (2b) Spectrum obtained with the \halpha\ grating during Visit 6.
The line visible in Knots C and D blueward of the [SII] doublet is He~I
$\lambda$6678. (2c) Spectrum obtained with the \halpha\ grating during Visit 7.}
\label{fig2a}
\end{figure}



\begin{figure}
\epsscale{.5}
\plotone{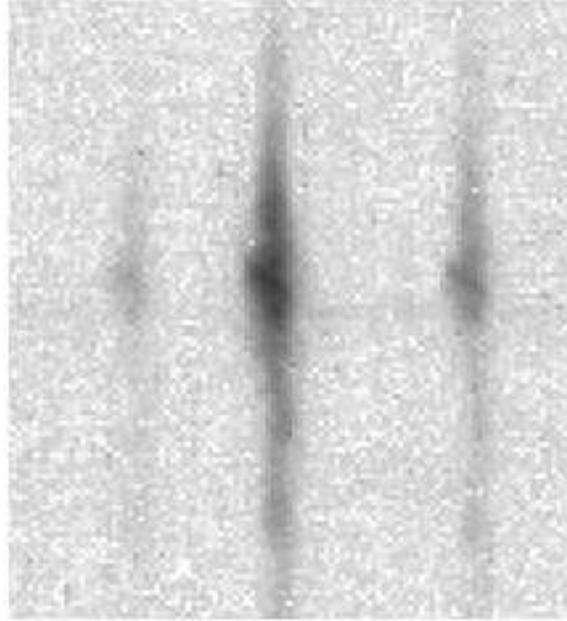}
\caption{Blowup of
spectrum of Knot B, including an unusual steep-gradient feature.}
\label{fig3}
\end{figure}

\begin{figure}
\centering
\includegraphics[angle=-90,scale=.5]{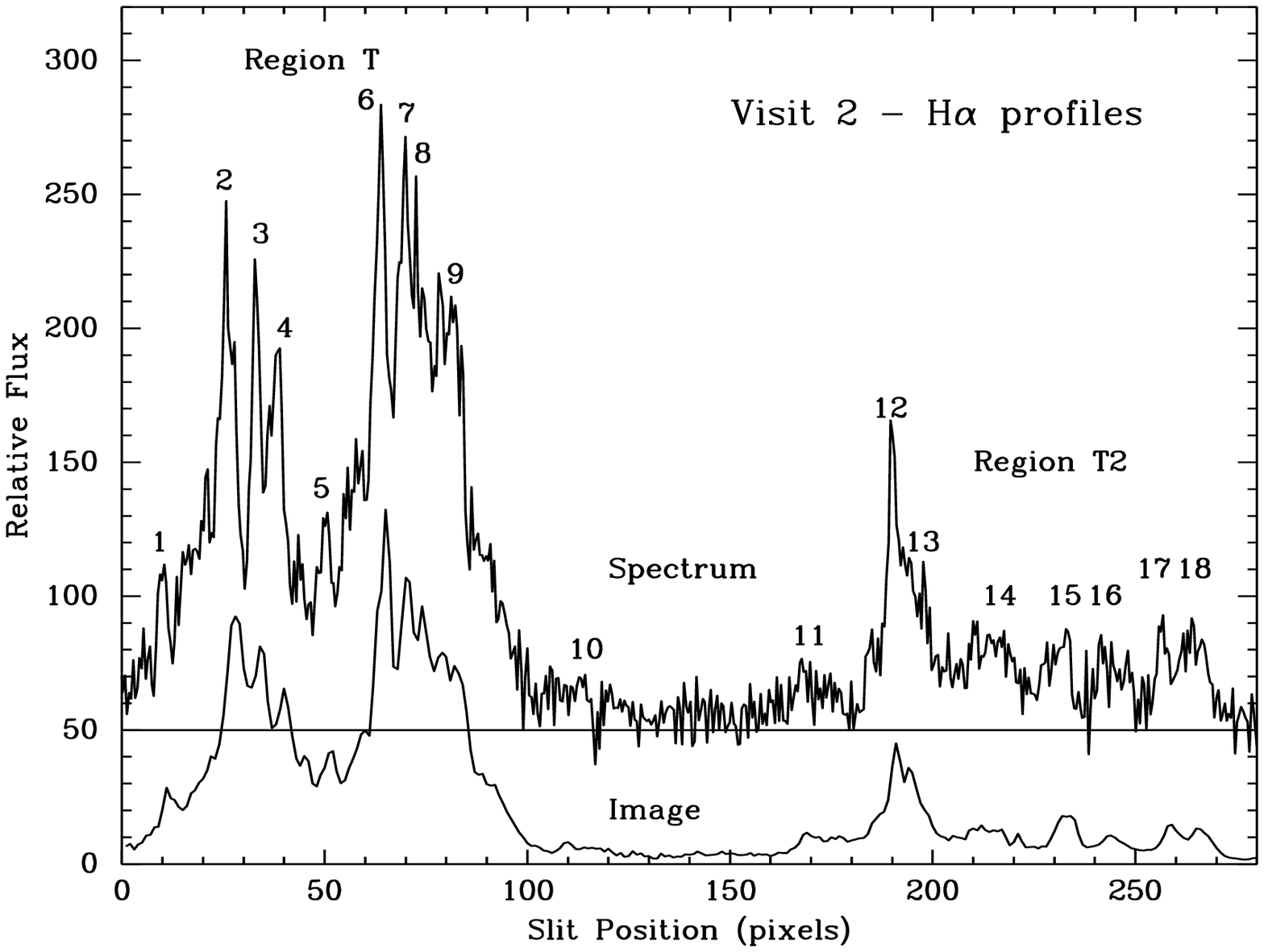}\\
\includegraphics[angle=-90,scale=.5]{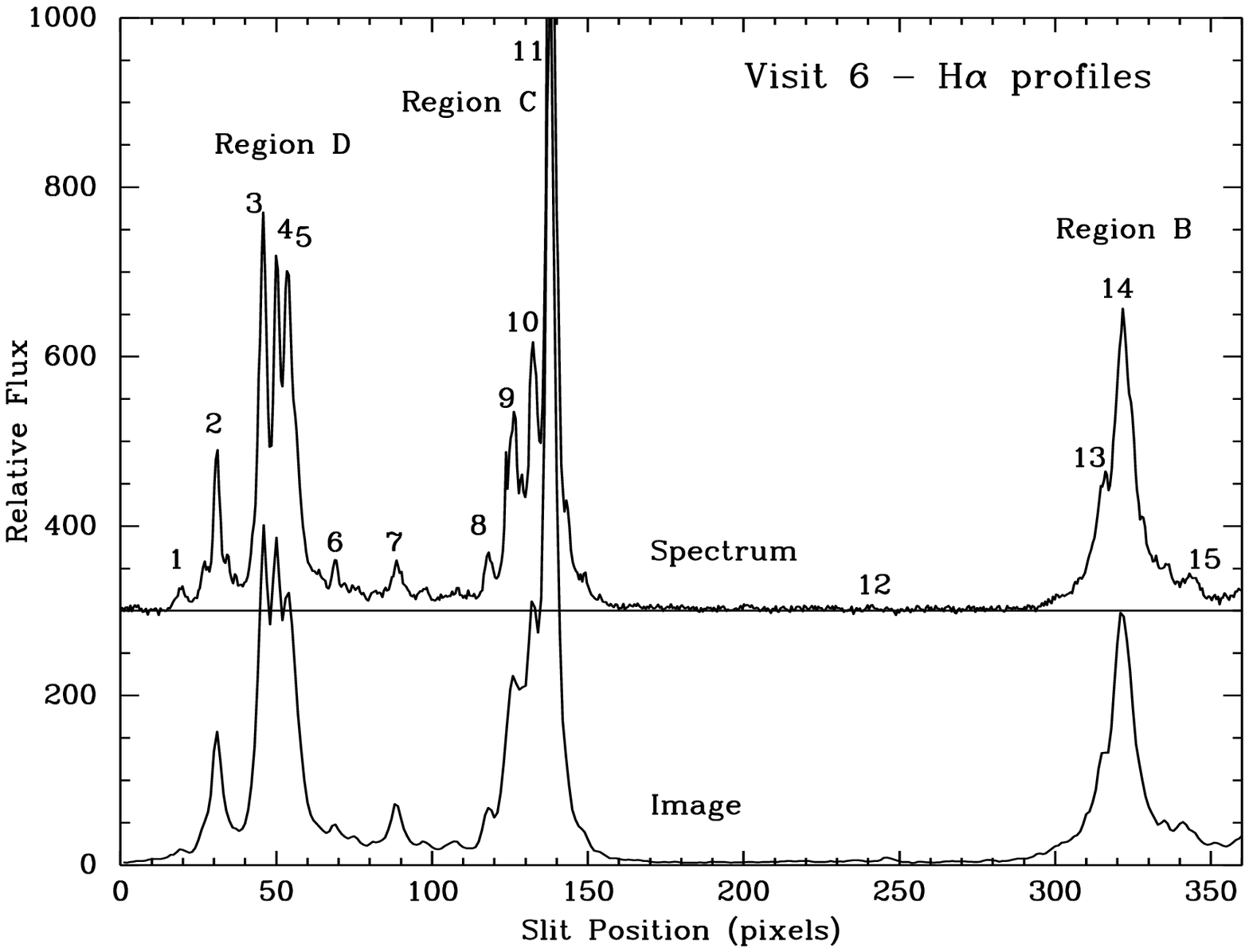}~
\includegraphics[angle=-90,scale=.5]{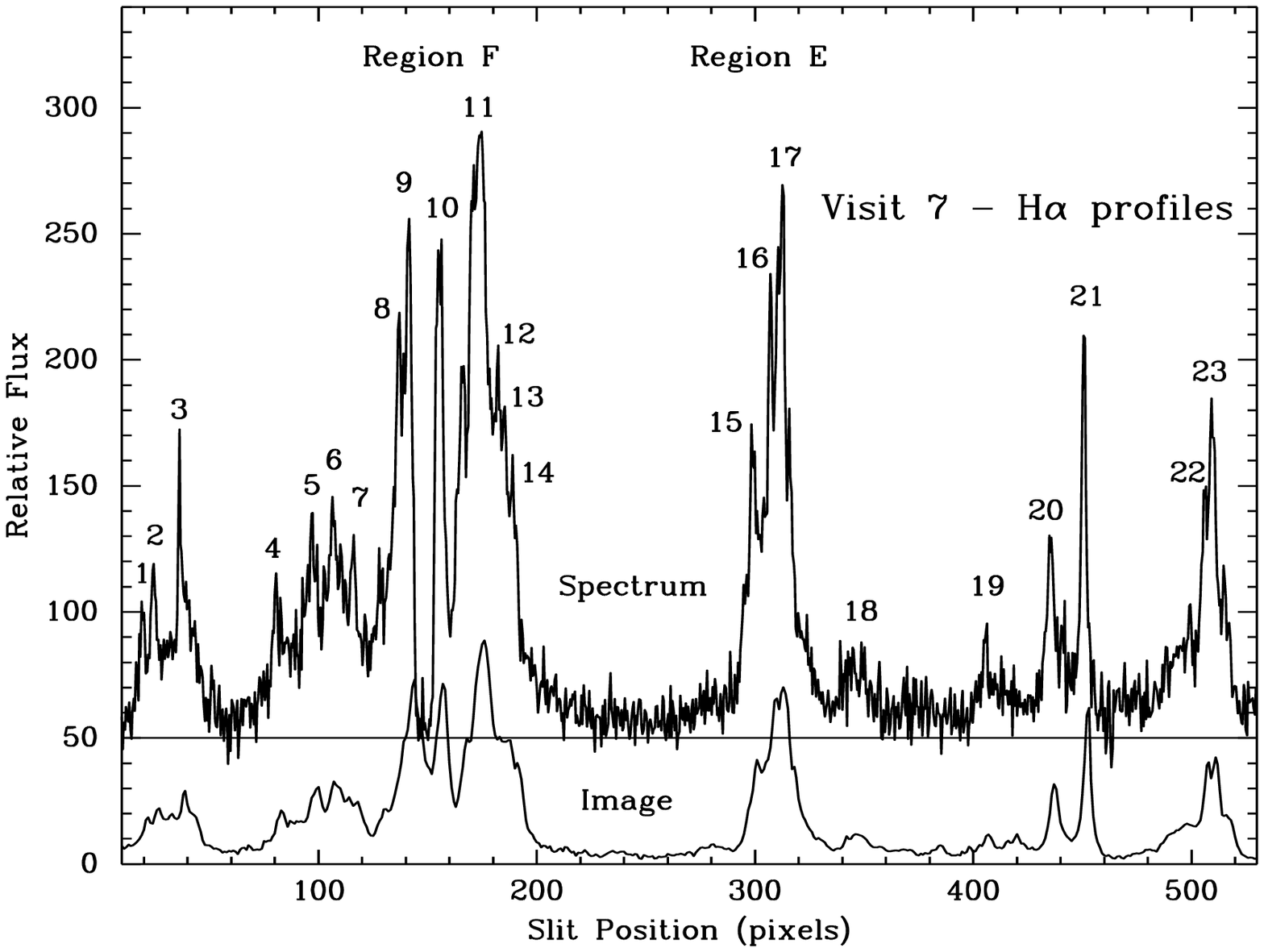}
\caption{(4a)  Comparison of H$\alpha$ profile along the slit during Visit 2
with an extraction from a simulated slit on the WFPC2 \halpha\ image.
ID numbers refer to objects in Table 1 and Figure 5a. (4b)  Comparison of H$\alpha$ profile along the slit during Visit 6
with an extraction from a simulated slit on the WFPC2 \halpha\ image.
ID numbers refer to objects in Table 1 and Figure 5b. (4c)  Comparison of H$\alpha$ profile along the slit during Visit 7
with an extraction from a simulated slit on the WFPC2 \halpha\ image.
ID numbers refer to objects in Table 1 and Figure 5c.}
\label{fig4a}
\end{figure}



\begin{figure}
\centering
\includegraphics[scale=.7]{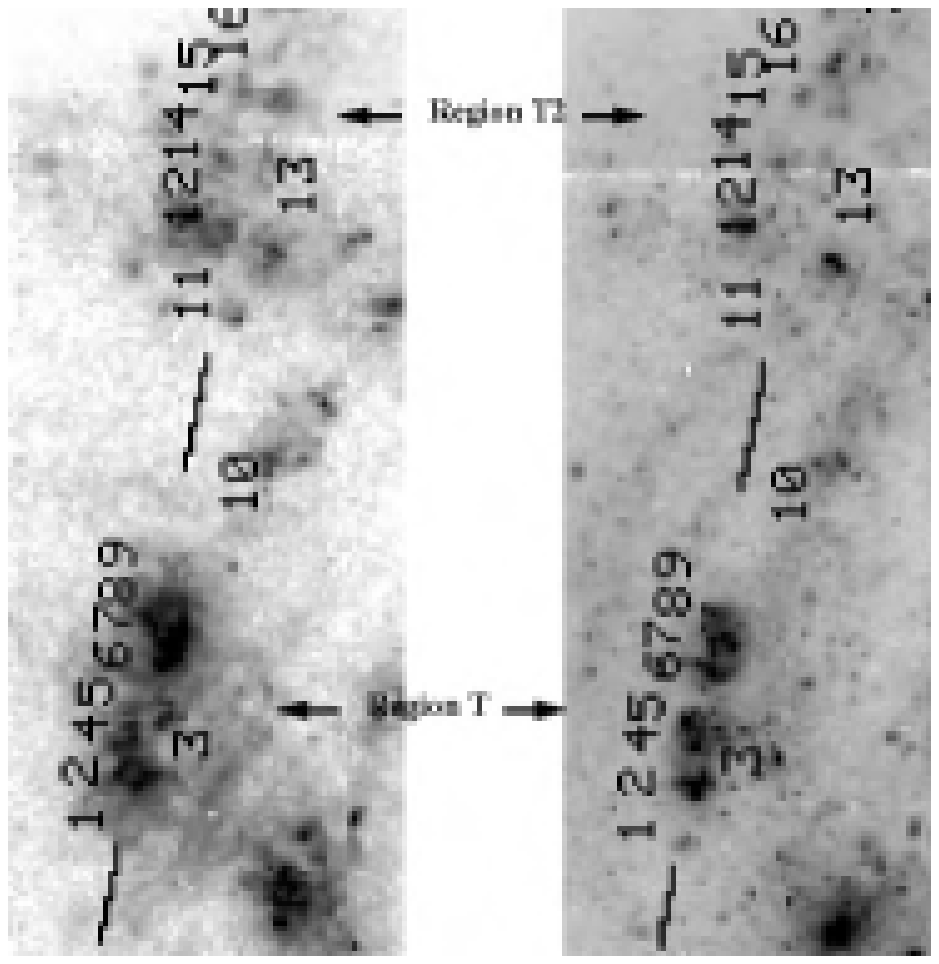}~~\includegraphics[scale=1]{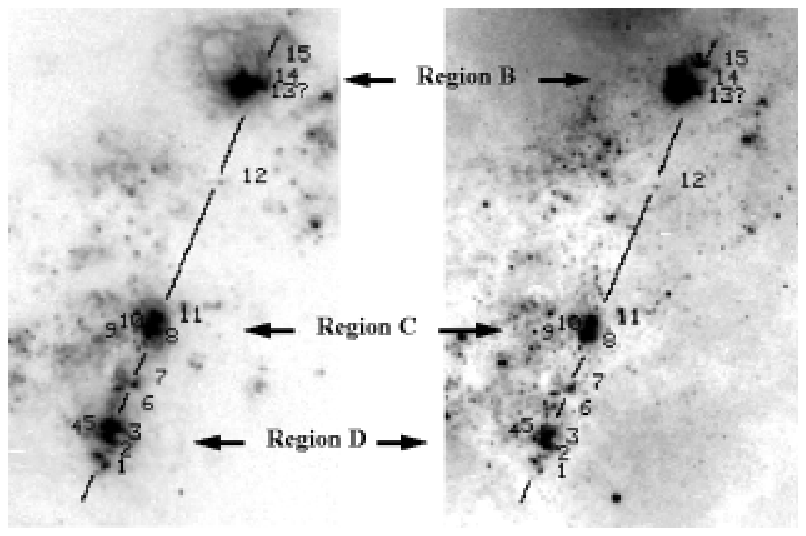}\\
\smallskip
\includegraphics[scale=1]{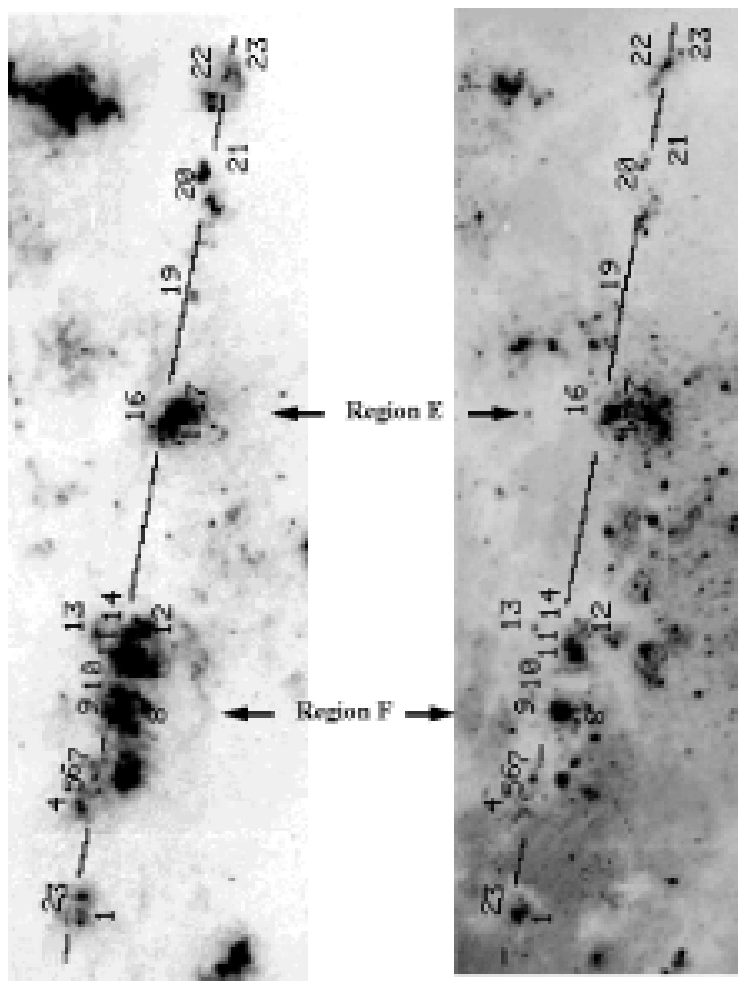}
\caption{(5a) Slit position and candidate clusters for Visit 2
superposed on the \halpha\ image. Note that the numbers refer to the candidate 
clusters defined in Table 1, and Figure 4. (5b) Slit position and candidate clusters for Visit 2
superposed on the $I$ image. (5c) Slit position and candidate clusters for Visit 6
superposed on the \halpha\ image (5d) Slit position and candidate clusters for Visit 6
superposed on the $I$ image (5e) Slit position and candidate clusters for Visit 7
superposed on the \halpha\ image (5f) Slit position and candidate clusters for Visit 7
superposed on the $I$ image.} 
\label{fig5a}
\end{figure}



\begin{figure}
\centering
\includegraphics[angle=-90,width=.7\textwidth]{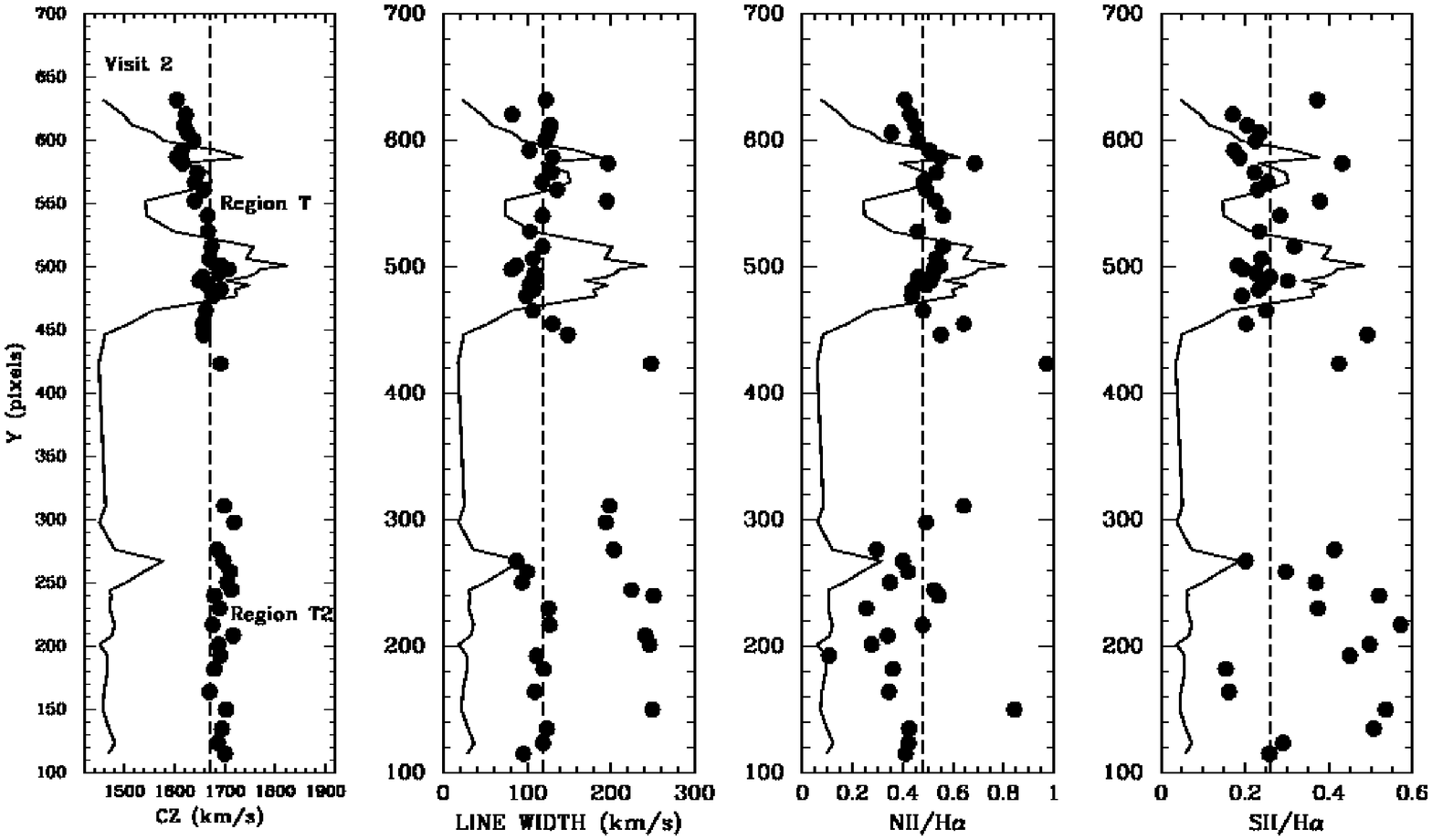}\\
\smallskip
\includegraphics[angle=-90,width=.7\textwidth]{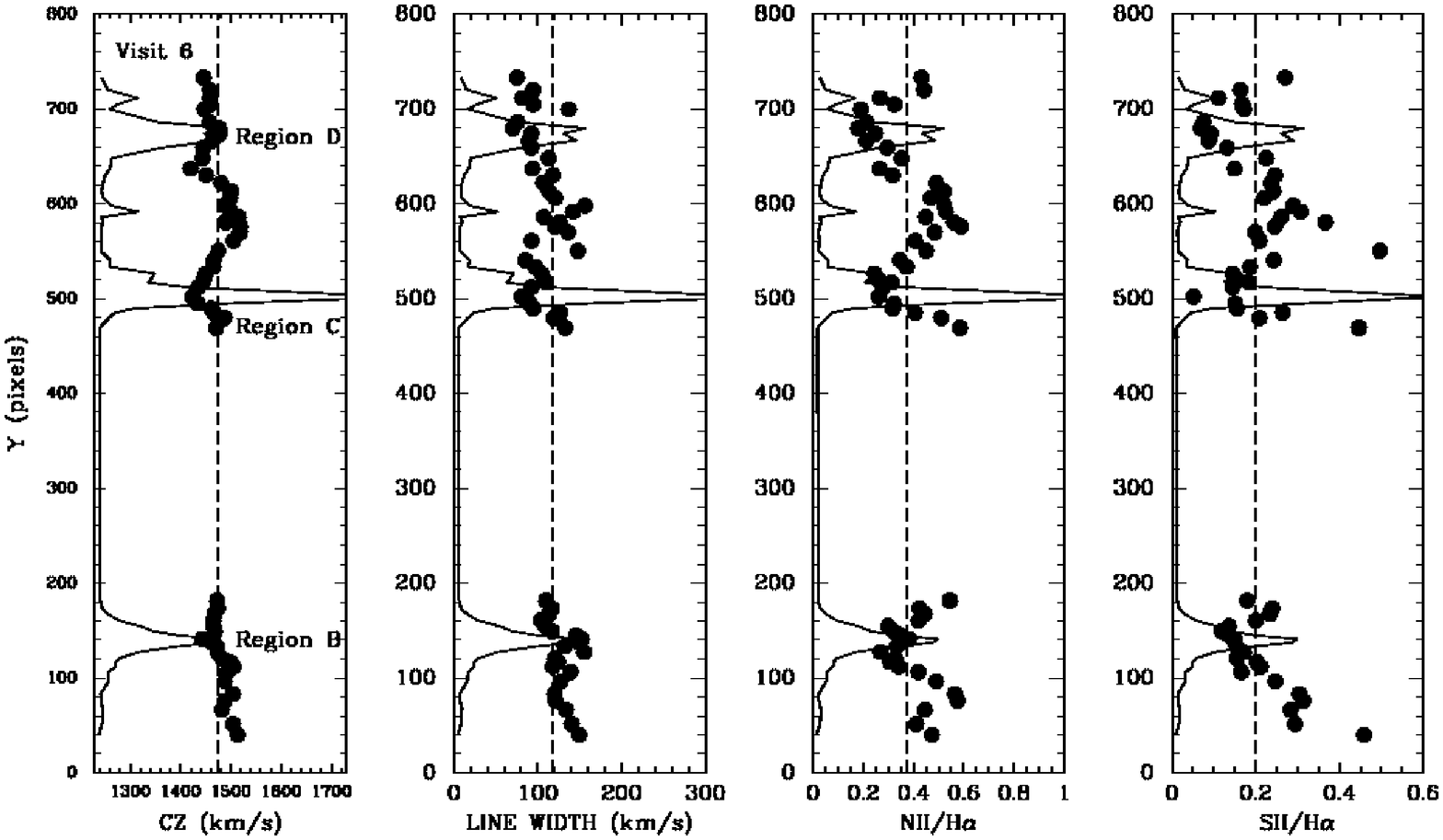}\\
\smallskip
\includegraphics[angle=-90,width=.7\textwidth]{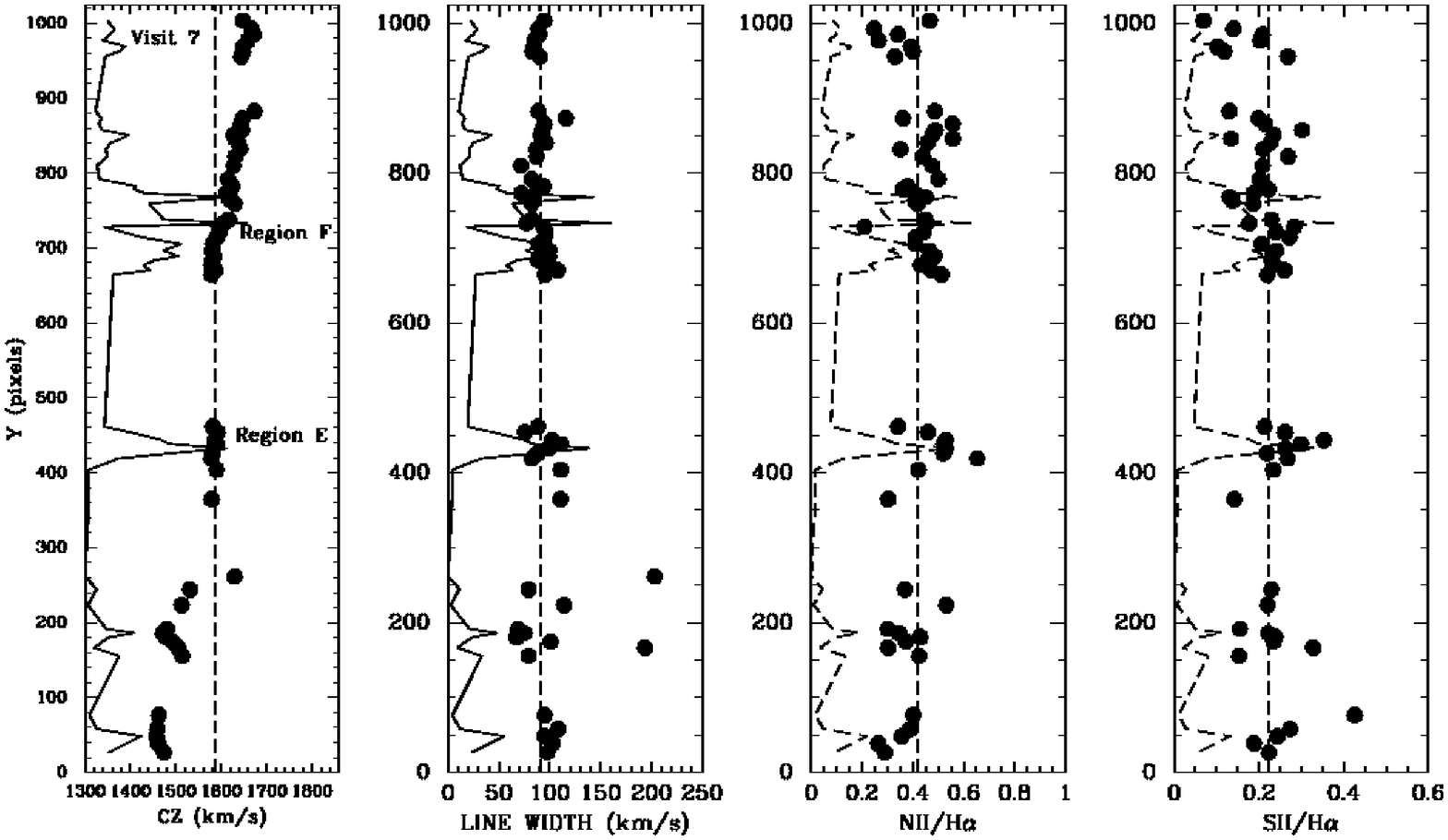}
\caption{(6a) Relative \halpha\ flux (solid lines), velocity $cz$,
\halpha\ line width, and [NII]/\halpha\ and [SII]/\halpha\ line ratios
for Visit 2. (6b) Relative \halpha\ flux (solid lines), velocity $cz$,
\halpha\ line width, and [NII]/\halpha\ and [SII]/\halpha\ line ratios
for Visit 6.(6c) Relative \halpha\ flux (solid lines), velocity $cz$,
\halpha\ line width, and [NII]/\halpha\ and [SII]/\halpha\ line ratios
for Visit 7.}
\label{fig6a}
\end{figure}




\begin{deluxetable}{ccccrcccccccc}
\tablecaption{Candidate Cluster Matches}
\tablewidth{0pt}
\tablehead{
& &  \colhead{Candidate} &  & & & & & \colhead{Line} & \colhead{Relative}  & \\
\colhead{Knot} &\colhead{Object $\#$} &\colhead{Cluster ID} &\colhead{Quality\tablenotemark{a}} &\colhead{M$_V$} &\colhead{log Age\tablenotemark{b}} &\colhead{log EW(H${\alpha}$)} &\colhead{Velocity} &\colhead{Width} &\colhead{H${\alpha}$ Flux} &\colhead{NII/H${\alpha}$} &\colhead{SII/H${\alpha}$}\\
&  &  & \colhead{(mag)} & \colhead{(yrs)} & & \colhead{(km/s)} & \colhead{(km/s)} & \colhead{(km/s)}\\[2pt]
\colhead{(1)} & \colhead{(2)} & \colhead{(3)} & \colhead{(4)} & \colhead{(5)} & \colhead{(6)} & \colhead{(7)} & \colhead{(8)} & \colhead{(9)} & \colhead{(10)} & \colhead{(11)} & \colhead{(12)}}
\startdata
\multicolumn{12}{c}{VISIT 2}\\ [2pt]
\hline
\noalign{\smallskip}
\colaa T \cola   \phn1\colb     7025\colc    3\cold  $-$8.1\cole    7.8\colf   --- \colg  1623\colh     82.1\coli      \phn61\colj     0.43\colk     0.17\eol
\colaa T \cola   \phn2\colb     6745\colc    3\cold  $-$11.0\cole    6.7\colf    2.2\colg  1614\colh    \llap{1}02.9\coli    212\colj      0.50\colk     0.17\eol
\colaa T \cola   \phn3\colb      --- \colc    3\cold  --- \cole   --- \colf   --- \colg  1646\colh    \llap{1}29.5\coli    198\colj     0.53\colk     0.22\eol
\colaa T \cola   \phn4\colb     6405\colc    3\cold $-$11.2\cole    6.7\colf    1.9\colg  1649\colh     \llap{1}27.0\coli     187\colj     0.48\colk     0.24\eol
\colaa T \cola   \phn5\colb     6206\colc    3\cold  $-$10.0\cole    6.8\colf    1.9\colg  1666\colh    \llap{1}18.6\coli     \phn99\colj     0.56\colk     0.28\eol
\colaa T \cola   \phn6\colb     5981\colc    2\cold $-$12.1\cole    6.8\colf    2.2\colg  1674\colh    \llap{1}18.6\coli    270\colj     0.56\colk     0.32\eol
\colaa T \cola   \phn7\colb     5784\colc    3\cold $-$11.6\cole    6.9\colf    2.1\colg  1681\colh     97.1\coli     290\colj     0.54\colk     0.21\eol
\colaa T \cola   \phn8\colb     5583\colc    3\cold $-$10.3\cole    6.8\colf    2.4\colg  1707\colh     81.3\coli    281\colj     0.52\colk      0.20\eol
\colaa T \cola   \phn9\colb      --- \colc  --- \cold  --- \cole   --- \colf   --- \colg  1666\colh    \llap{1}02.9\coli    262\colj     0.49\colk     0.24\eol
\colaa T \cola  10\colb      --- \colc  --- \cold  --- \cole   --- \colf   --- \colg   --- \colh     --- \coli     --- \colj     --- \colk     --- \eol
\colaa T2 \cola  11\colb     4118\colc    1\cold $-$11.6\cole     6.0\colf    3.7\colg  1699\colh    \llap{1}98.7\coli     \phn34\colj     0.64\colk     0.77\eol
\colaa T2 \cola  12\colb     3635\colc    2\cold  $-$10.0\cole    6.2\colf    3.4\colg  1696\colh     87.2\coli    126\colj      0.40\colk      0.20\eol
\colaa T2 \cola  13\colb      --- \colc  --- \cold  --- \cole   --- \colf   --- \colg   --- \colh     --- \coli     --- \colj     --- \colk     --- \eol
\colaa T2 \cola  14\colb     3357\colc    1\cold  $-$9.6\cole    6.5\colf    3.2\colg  1676\colh     \llap{1}27.0\coli     \phn48\colj     0.48\colk     0.57\eol
\colaa T2 \cola  15\colb     3172\colc    1\cold  $-$12.0\cole     6.0\colf    3.6\colg  1679\colh    \llap{1}20.1\coli      \phn38\colj     0.36\colk     0.15\eol
\colaa T2 \cola  16\colb     3118\colc    2\cold $-$12.9\cole    6.5\colf    2.8\colg  1670\colh    \llap{1}09.1\coli     \phn32\colj     0.34\colk     0.16\eol
\colaa T2 \cola  17\colb     2993\colc    1\cold $-$11.6\cole    6.5\colf     3.0\colg  1694\colh    \llap{1}23.8\coli     \phn39\colj     0.43\colk     0.51\eol
\colaa T2 \cola  18\colb     2972\colc    1\cold $-$11.8\cole     6.0\colf    3.7\colg  1686\colh    \llap{1}19.1\coli      \phn49\colj     0.42\colk     0.29\eol
\cutinhead{VISIT 6}\\
\colaa D \cola   \phn1\colb     2578\colc    1\cold   $-$9.0\cole     6.0\colf    3.5\colg  1446\colh     75.4\coli     \phn79\colj     0.43\colk     0.27\eol
\colaa D \cola   \phn2\colb     2511\colc    1\cold $-$10.8\cole    6.5\colf    3.5\colg  1458\colh     81.6\coli    503\colj     0.27\colk     0.11\eol
\colaa D \cola   \phn3\colb     2428\colc    2\cold $-$11.6\cole    6.6\colf    3.4\colg  1477\colh      70.\phn\coli   \llap{1}568\colj     0.18\colk     0.07\eol
\colaa D \cola   \phn4\colb     2410\colc    2\cold $-$12.7\cole    6.8\colf    2.5\colg  1479\colh     92.5\coli   \llap{1}304\colj     0.25\colk     0.09\eol
\colaa D \cola   \phn5\colb     2379\colc    3\cold $-$11.4\cole     6.0\colf    3.5\colg  1464\colh     89.1\coli   \llap{1}458\colj     0.21\colk     0.09\eol
\colaa D \cola   \phn6\colb     2314\colc    2\cold  --- \cole   --- \colf    3.4\colg  1420\colh     93.8\coli    177\colj     0.27\colk     0.15\eol
\colaa \cola   \phn7\colb     2202\colc    1\cold $-$11.8\cole     6.0\colf    3.6\colg  1489\colh    \llap{1}56.2\coli    198\colj     0.52\colk     0.29\eol
\colaa C \cola   \phn8\colb     2096\colc    1\cold  $-$11.0\cole     6.0\colf    3.6\colg  1466\colh     85.3\coli    185\colj     0.35\colk     0.24\eol
\colaa C \cola   \phn9\colb      --- \colc    3\cold  --- \cole   --- \colf   --- \colg  1445\colh    \llap{1}09.7\coli    626\colj     0.31\colk     0.18\eol
\colaa C \cola  10\colb     2028\colc    3\cold $-$12.4\cole    6.5\colf    3.2\colg  1436\colh     92.7\coli   \llap{1}102\colj     0.28\colk     0.14\eol
\colaa C \cola  11\colb     2002\colc    2\cold $-$13.3\cole     6.0\colf    3.5\colg  1424\colh     80.9\coli   \llap{3}629\colj     0.26\colk     0.05\eol
\colaa \cola  12\colb     1496\colc    1\cold  $-$13.0\cole     6.0\colf    3.7\colg   --- \colh     --- \coli     --- \colj     --- \colk     --- \eol
\colaa B \cola  13\colb      --- \colc    3\cold  --- \cole   --- \colf   --- \colg  1469\colh    \llap{1}16.8\coli     686\colj     0.32\colk     0.12\eol
\colaa B \cola  14\colb     1233\colc    2\cold $-$12.9\cole    6.5\colf    3.1\colg  1444\colh    \llap{1}52.1\coli   \llap{1}490\colj     0.38\colk     0.15\eol
\colaa B \cola  15\colb     1143\colc    3\cold  $-$12.0\cole    6.7\colf    2.1\colg  1506\colh    \llap{1}17.9\coli    241\colj     0.34\colk     0.21\eol
\cutinhead{VISIT 7}\\
\colaa \cola   \phn1\colb     9838\colc    3\cold  $-$8.6\cole    6.9\colf    2.4\colg  1648\colh     94.7\coli      \phn61\colj     0.47\colk     0.07\eol
\colaa \cola   \phn2\colb     9806\colc    3\cold  $-$11.\phn\cole    6.8\colf    1.6\colg  1667\colh     88.6\coli     \phn81\colj     0.25\colk     0.14\eol
\colaa \cola   \phn3\colb     9745\colc    1\cold  $-$9.9\cole     6.0\colf    3.5\colg  1649\colh     83.9\coli    110\colj     0.39\colk      0.10\eol
\colaa F \cola   \phn4\colb     9196\colc    3\cold $-$11.2\cole     6.0\colf    3.7\colg  1674\colh     88.8\coli     \phn29\colj     0.49\colk     0.13\eol
\colaa F \cola   \phn5\colb     9152\colc    1\cold  $-$8.3\cole     6.0\colf    3.4\colg  1628\colh      91.0\coli    118\colj     0.48\colk     0.23\eol
\colaa F \cola   \phn6\colb     8996\colc    1\cold $-$11.9\cole    6.8\colf    1.8\colg  1643\colh     86.7\coli     \phn58\colj     0.35\colk     0.21\eol
\colaa F \cola   \phn7\colb      --- \colc    3\cold  --- \cole   --- \colf   --- \colg  1627\colh     71.6\coli     \phn33\colj     0.48\colk     0.21\eol
\colaa F \cola   \phn8\colb     8437\colc    2\cold $-$10.2\cole     6.0\colf    3.5\colg  1610\colh     71.9\coli    164\colj      0.40\colk     0.19\eol
\colaa F \cola   \phn9\colb     8306\colc    3\cold $-$10.7\cole     6.0\colf    3.5\colg  1622\colh     82.3\coli    278\colj     0.41\colk     0.14\eol
\colaa F \cola  10\colb     8109\colc    3\cold  $-$9.8\cole     6.0\colf    3.5\colg  1617\colh     82.5\coli    216\colj     0.45\colk     0.23\eol
\colaa F \cola  11\colb      --- \colc    3\cold  --- \cole   --- \colf   --- \colg  1580\colh     99.5\coli    215\colj     0.46\colk     0.24\eol
\colaa F \cola  12\colb     7488\colc    2\cold  $-$11.\phn\cole    6.4\colf    3.3\colg  1580\colh     88.2\coli    189\colj     0.46\colk     0.23\eol
\colaa F \cola  13\colb    \phn\phn\phn0\colc    3\cold  --- \cole   --- \colf   --- \colg  1579\colh     95.2\coli    161\colj     0.43\colk     0.24\eol
\colaa F \cola  14\colb     7391\colc    2\cold $-$11.4\cole     6.0\colf    3.7\colg  1578\colh     95.3\coli     \phn76\colj     0.51\colk     0.22\eol
\colaa E \cola  15\colb     5254\colc    2\cold $-$10.6\cole    6.6\colf    3.2\colg  1591\colh     75.5\coli    124\colj     0.46\colk     0.26\eol
\colaa E \cola  16\colb     5028\colc    2\cold $-$12.1\cole    6.2\colf    3.4\colg  1591\colh    \llap{1}10.8\coli    232\colj     0.52\colk      0.30\eol
\colaa E \cola  17\colb     5023\colc    2\cold $-$11.3\cole    6.9\colf    2.5\colg  1582\colh     87.8\coli     239\colj     0.52\colk     0.22\eol
\colaa E \cola  18\colb      --- \colc    3\cold  --- \cole   --- \colf   --- \colg  1579\colh    \llap{1}10.6\coli     \phn11\colj      0.30\colk     0.14\eol
\colaa \cola  19\colb     3884\colc    1\cold $-$8.4\cole    6.1\colf    3.4\colg  1531\colh     79.1\coli     \phn32\colj     0.37\colk     0.23\eol
\colaa \cola  20\colb     3193\colc    1\cold $-$10.9\cole     6.0\colf    3.6\colg  1472\colh     74.8\coli    \llap{1}334\colj     0.35\colk     0.22\eol
\colaa \cola  21\colb     3050\colc    1\cold $-$10.8\cole     6.0\colf    3.5\colg  1514\colh     79.1\coli     \phn93\colj     0.43\colk     0.15\eol
\colaa \cola  22\colb     2666\colc    1\cold $-$11.3\cole    6.9\colf     2.0\colg  1457\colh     94.9\coli    154\colj     0.36\colk     0.24\eol
\colaa \cola  23\colb     2612\colc    2\cold  $-$7.4\cole    6.8\colf    2.5\colg  1474\colh     97.6\coli     \phn64\colj     0.29\colk     0.22\eol
\enddata
\tablenotetext{a}{1 = Good match between H${\alpha}$ object and cluster;
2 = possible match, 3 = poor match (i.e., H${\alpha}$ diffuse or shell
like). Note that in cases when the emission appears to be diffuse, we
include nearby clusters in the table in order to obtain information
about the physical characteristics within the knot.}

\tablenotetext{b}{Ages based on  a slightly modified version of the technique used
by Whitmore \& Zhang (2002) (i.e., using Bruzual\& Charlot 2003 models). }
\end{deluxetable}

\begin{deluxetable}{ccccccccc}
\tablewidth{0pt}
\tablecaption{Velocity Dispersions and Mean Values of Various Parameters for Selected Knots} 
\tablehead{
&\colhead{Number of} &             &\colhead{\halpha\ Velocity} &         & \colhead{\halpha\ Line} &         &        &\colhead{HI Velocity}\\
\colhead{Knot} &\colhead{Clusters\tablenotemark{a}}  &\colhead{Quality\tablenotemark{b}} &\colhead{Dispersions} 
&\colhead{log Age\tablenotemark{c}} &\colhead{Width} &\colhead{[NII]/H$\alpha$} &\colhead{[SII]/H$\alpha$} 
&\colhead{Dispersion\tablenotemark{d}}\\
&           &             &\colhead{(\kms)} &\colhead{(yr)} &\colhead{(\kms)} &         &        &\colhead{(\kms)}}
\startdata
T  &  10 (7$''$) & 2.9 & 14.0 & 6.8 & 107 (6) & 0.51 (.02) & 0.23 (.02)  & 15--30  \\
T2 &   \phn7 (8$''$) & 1.3 &  \phn9.8 & 6.3 & 114 (6) & 0.44 (.04) & ---        & $\sim$15 \\
B  &   \phn3 (2$''$) & 2.5 &  \phn3.8 & 6.6 & 117 (1) & 0.36 (.02) & 0.18 (.03) & $\sim$15 \\
C  &   \phn4 (4$''$) & 2.0 & 10.5 & 6.2 &  \phn92 (6) & 0.30 (.03) & 0.14 (.06) & $\sim$15 \\
D  &   \phn6 (2$''$) & 1.8 & 15.6 & 6.4 &  \phn84 (4) & 0.27 (.04) & 0.13 (.04) & ---        \\
E  &   \phn4 (2$''$) & 2.0 &  \phn3.1 & 6.5 &  \phn96 (4) & 0.50 (.02) & 0.26 (.02) & $\sim$30 \\
F  &  11 (10$''$) & 2.1 & 10.5 & 6.2 &  \phn87 (3) & 0.44 (.05) & 0.20 (.01) & $\sim$30 \\
Mean& 6.4 (5$''$) & 2.1 &  \phn9.6 & 6.4\rlap{3}& 100 \hphantom{(8)} & 0.40 \hphantom {(.02)} & 0.19 \hphantom {(.02)} &
\enddata
\tablerefs{Numbers in parentheses are standard deviations of the mean.}
\tablenotetext{a}{Numbers in parenthesis are the spatial extent of the clusters in the Knot.}
\tablenotetext{b}{From Table 1.}
\tablenotetext{c}{From Whitmore \& Zhang (2004).}
\tablenotetext{d}{From Hibbard et al. (2001).}
\end{deluxetable}

\end{document}